\documentclass[10pt,
	nofootinbib,
         amsmath,
         amssymb,
         superscriptaddress,
         aps,
         prb, 
         nolongbibliography,
         floatfix,
         twocolumn
]{revtex4-2}
\usepackage[english]{babel}
\usepackage[utf8]{inputenc} 
\usepackage{dsfont}
\usepackage{siunitx}
\usepackage[caption=false,label font=bf,labelformat=simple]{subfig}
\usepackage{graphicx}     
\usepackage{color}
\usepackage[normalem]{ulem}
\usepackage{soul}
\usepackage{cancel}
\usepackage{hyperref}
\definecolor{Jerome}{rgb}{0.8, 0.0, 0.3}

\definecolor{Noe}{rgb}{0.,0.8,0.3}

\begin{document}
\title{Anyon braiding on the single edge of a fractional quantum Hall state}
\author{Flavio Ronetti}
\affiliation{Aix Marseille Univ, Universit\'e de Toulon, CNRS, CPT, Marseille, France}
\author{Noé Demazure}
\affiliation{Aix Marseille Univ, Universit\'e de Toulon, CNRS, CPT, Marseille, France}
\author{Jérôme Rech}
\affiliation{Aix Marseille Univ, Universit\'e de Toulon, CNRS, CPT, Marseille, France}
\author{Thibaut Jonckheere}
\affiliation{Aix Marseille Univ, Universit\'e de Toulon, CNRS, CPT, Marseille, France}
\author{Benoît Grémaud}
\affiliation{Aix Marseille Univ, Universit\'e de Toulon, CNRS, CPT, Marseille, France}
\author{Laurent Raymond}
\affiliation{Aix Marseille Univ, Universit\'e de Toulon, CNRS, CPT, Marseille, France}
\author{Masayuki Hashisaka}
\affiliation{Institute for Solid State Physics, University of Tokyo, 5-1-5 Kashiwanoha, Kashiwa, Japan}
\author{Takeo Kato}
\affiliation{Institute for Solid State Physics, University of Tokyo, 5-1-5 Kashiwanoha, Kashiwa, Japan}
\author{Thierry Martin}
\affiliation{Aix Marseille Univ, Universit\'e de Toulon, CNRS, CPT, Marseille, France}

\begin{abstract}
Anyons are quasiparticles with fractional statistics, bridging between fermions and bosons. We propose an experimental setup to measure the statistical angle of topological anyons emitted from a quantum point contact (QPC) source. The setup involves an $\Omega$-shaped junction along a fractional quantum Hall liquid edge, formed by defining a droplet with two negatively biased gates. In the weak tunneling regime, we calculate the charge current, showing its time evolution depends solely on the anyons’ statistical properties, with temperature and scaling dimension affecting only the constant prefactor. We compute the cross-correlation between the anyon current transmitted from the source and the current after the $\Omega$-junction, providing a direct method to detect anyon braiding statistics.
\end{abstract}
\maketitle

\textit{Introduction:-} Anyons are quasiparticles that emerge in two-dimensional systems and obey statistical rules that differ from the familiar fermionic and bosonic statistics~\cite{Leinaas77,Wilczek82,Arovas84}. In the fractional quantum Hall effect (FQHE), they appear as quasiparticles carrying fractional charge and exhibiting fractional statistics~\cite{Tsui82,Laughlin83}. The fractional value of charge has been accessed for anyons a long time ago by measuring the shot noise in the state $\nu = 1/3$~\cite{Kane94,Saminadayar97,dePicciotto97,Crepieux04,Martin05} and further confirmed recently by other types of measurement~\cite{JensMartin04,Kapfer19,Bisognin19}. Concerning the fractional statistics, more complex experiments have been proposed to extract the braiding properties of anyons~\cite{Chamon97,Hou97,Safi01,Vishveshwara03,Bishara08,Halperin11,Campagnano12,Rosenow12,Levkivskyi12,Lee19,Carrega21}. Recently, experimental breakthroughs have provided compelling evidence also for the fractional statistics of anyons in FQHE systems at filling factor $\nu=1/3$ in the Fabry-Perot~\cite{Nakamura20} and collider~\cite{Bartolomei20} configurations. These first groundbreaking results have been subsequently confirmed and extended to other filling factors by additional experimental evidence~\cite{Ruelle23,Kundu23,Lee23,Glidic23,Nakamura23,Willett23,Werkmeister24,Werkmeister25}.

The synergy between numerous theoretical and experimental studies has firmly established that the primary mechanism underlying the physics of the anyon collider is the so-called time-domain braiding~\cite{Rosenow16,Han16,Lee22,Morel22,Mora22,Schiller23,Jonckheere23}. In this configuration, anyons are randomly emitted by a source quantum point contact (QPC) tuned to the weak-backscattering regime~\cite{Kane94}. This dilute stream of anyons is sent to a central QPC, also in the weak-backscattering regime, at which quasiparticle/quasihole pairs can be spontaneously produced. An anyon emitted by the source QPC can impinge at the central QPC before or after the spontaneous excitation of a pair, thus giving rise to a non-trivial quantum interference between two events with a different statistical phase. The latter depends on the statistical angle $\pi \lambda$ and, being a topologically protected quantity, is independent of the sample details. Nevertheless, the probability of the spontaneous creation of pairs scales as a temperature power-law whose exponent is controlled by the scaling dimension $\delta$. Importantly, the scaling dimension is not a topologically protected quantity and can be renormalized due to effects related to the specific sample~\cite{Braggio12}. Indeed, despite the theoretical effort to propose alternatives for assessing the scaling dimension of anyons in the FQHE~\cite{Snizhko15,Rech20,Ebisu22,Schiller22,Bertin23,Iyer23,Acciai24}, experiments still cannot provide a conclusive answer~\cite{Veillon24,Schiller24,Ruelle24,Ramon25}. 

\begin{figure}[t]
\centering
	\includegraphics[width=0.7\linewidth]{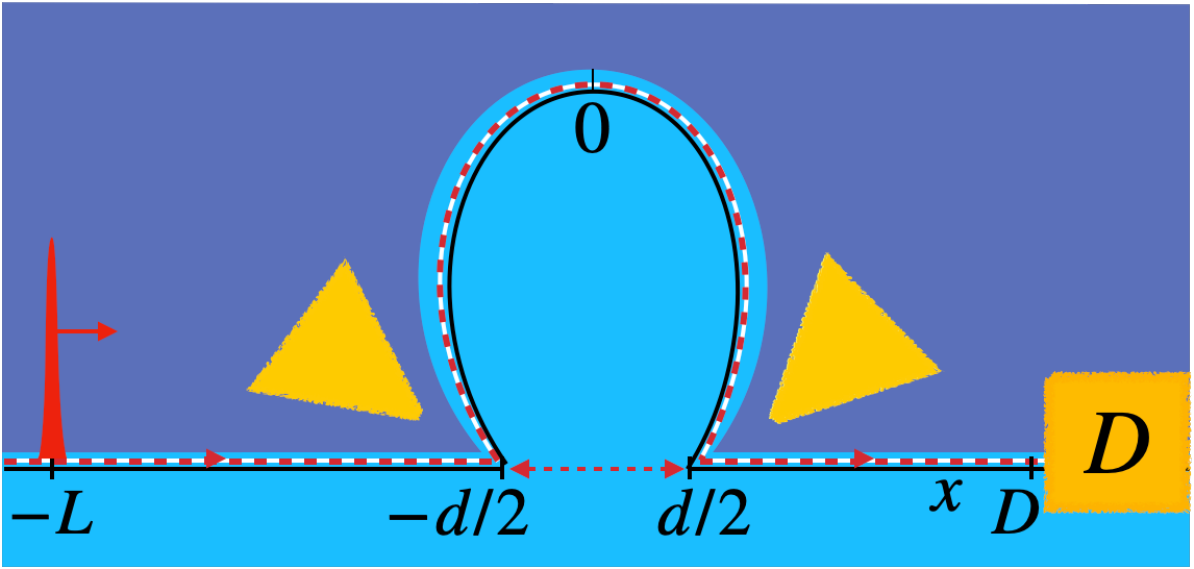}
	\caption{Sketch of the setup: we consider a single edge of a fractional quantum Hall bar. By means of negatively charged voltage gates, a droplet of total length $d$ is formed. We term this configuration ``$\Omega-$junction". By choosing a curvilinear abscissa $x$ along the edge and setting $x=0$ at the center of the droplet, the tunneling occurs between $x=-d/2$ and $x=d/2$. We focus on the weak-tunneling regime where fractional quasiparticles are tunneling through the junction. Anyons are assumed to be injected at $x=-L$ and the charge current is detected at the drain $x=D$.}
	\label{fig:scheme}
\end{figure}

In this Letter, we propose a new experimental setup to measure the statistical angle $\pi \lambda$ of topological anyons emitted by a source QPC without any need to extract independently the scaling dimension $\delta$. We consider the single edge of a FQHE liquid at $\nu=1/\left({2n+1}\right)$, with $n \in \mathbb{N}$, in the presence of an $\Omega-$shaped junction (see Fig.~\ref{fig:scheme}). The latter is obtained by geometrically defining a droplet in the fractional quantum Hall fluid. This configuration is reminiscent of experiments performed at integer filling factor in the context of the mesoscopic capacitor~\cite{Gabelli07,Parmentier12}. We compute the time-dependent charge current for isolated delta-like anyon pulses. We show that the time evolution of charge current is controlled by the exchange statistics of anyons, while other non-universal factors, such as temperature and scaling dimension, just renormalize the current prefactor. Then, we turn our attention to a dilute stream of anyons emitted by a source QPC placed upstream with respect to the $\Omega-$junction, following the concept of the anyon collider configuration. We investigate finite-frequency current cross-correlations at two system outputs, showing how they can detect anyon braiding and directly extract the statistical angle $\pi \lambda$. Our proposal allows for the extraction of the statistical angle independently of the scaling dimension, similar to recent experimental protocols~\cite{Ruelle24}. This key result could pave the way for future experiments in anyon manipulation within the FQHE, contributing to the topologically protected control of quantum information.

\textit{Model:-}
We consider the edge of a quantum Hall bar tuned into the fractional regime at filling factor $\nu = 1/(2n+1)$, with $n \in \mathbb{N}$. Belonging to the Laughlin sequence, a single channel is present on the edge: the corresponding quasiparticles have a charge $e^* = e \nu$ and a statistical angle $ \pi \lambda = \pi \nu$.

The Hamiltonian describing the single channel on one edge is a bosonic conformal field theory (CFT)~\cite{Wen95}
\begin{equation}
	\label{eq:HamEdge}
	H_0 = \frac{u}{4\pi} \int dx~\left[\partial_x\phi(x)\right]^2,
\end{equation} 
where $u$ is the velocity of the chiral boson field $\phi(x,t) = \phi(x-ut,0) = \phi(0,t-x/u)$.  The description in terms of a CFT naturally arises for one dimensional gapless liquids. In addition, the chiral boson field satisfies the commutation relation $\left[\phi(x,t),\phi(x',t')\right] = i \pi \text{sign}\left[x-x'-u\left(t-t'\right)\right]$.

In order to investigate the braiding of anyons, we deform this single edge state such that a droplet of total perimeter $d>0$ is created. As a result, an $\Omega-$shaped junction is formed as presented in Fig.~\ref{fig:scheme}. The coordinate $x$ is the curvilinear abscissa along the edge and $x=0$ is chosen at the center of the loop (see Fig.~\ref{fig:scheme}), such that the tunneling occurs between $x=-d/2$ and $x=d/2$. We refer to the latter point as the \textit{junction} and to the region between $-d/2<x<d/2$ as the \textit{loop}. The corresponding tunneling Hamiltonian is~\cite{Kane94}
\begin{equation}
	\label{eq:HamTun}
	H_T = \Lambda e^{i\kappa} \Psi^{\dagger}(d/2)\Psi(-d/2) + \text{H.c.},
\end{equation}
where $\Lambda$ is a real tunneling amplitude, $\kappa$ is the complex phase of the tunneling amplitude. By means of negatively biased metallic gates, the tunneling amplitude can be tuned: we focus on the weak-tunneling limit. In Eq.~\eqref{eq:HamTun} we introduced the quasi-particle creation and destruction operators connected to the boson field of the edge through the bosonization identity~\cite{vonDelft98}
\begin{equation}
	\Psi(x) = \frac{\mathcal{F}}{\sqrt{2\pi a}} e^{-i \sqrt{\nu}\phi(x)},
\end{equation}
where $\mathcal{F}$ is the Klein factor and $a$ is a short-length cut-off. 

\textit{Correlation functions:-} It is instructive to present the correlation functions for the quasiparticle vertex operators in order to introduce the concepts of scaling dimensions and statistical angle. The correlation function for the quasi-particle operators is given by 
\begin{align}
    \left\langle T_K \left[e^{i \sqrt{\nu} \phi\left(x,t \eta\right)} \right.\right. & \left. \left.  e^{-i \sqrt{\nu} \phi\left(x',t' \eta'\right)} \right]\right\rangle = e^{\delta \mathcal{G}\left(t-t'-\frac{x-x'}{u}\right)} \nonumber\\
    & \qquad \times e^{-\frac{i \lambda\pi }{2}\sigma_{tt'}^{\eta\eta'}\text{sign}\left(t-t' - \frac{x-x'}{u}\right)}\label{eq:CorrelationAnyons}
\end{align}
where $\mathcal{G}\left(t-t'-\left(x-x'\right)/{u}\right)\equiv\frac{1}{2}\left\langle \left\{\phi\left(x,t\right),\phi\left(x',t'\right)\right\}\right\rangle - \left\langle\phi^2\left(0,0\right)\right\rangle$ contains all the information about the correlation of boson fields and $\sigma_{tt'}^{\eta\eta'} = \text{sign}\left(t-t'\right)\left(\eta+\eta'\right)/2 +  \left(\eta'-\eta\right)/2$ ensures the correct ordering along the Keldysh contour. In the above expression we also introduced the scaling dimension $\delta$ and the statistical angle $\lambda$. The first term becomes independent of the Keldysh indices because of the definition of the anti-commutator itself and its explicit form for a free bosonic theory is given by
\begin{equation}
\mathcal{G}\left(t\right) = -\ln \left|\frac{\sinh\left(i\frac{\pi}{\beta}\tau_0\right)}{\sinh\left[\frac{\pi}{\beta}\left(t+i\tau_0\right)\right]}\right|,
\end{equation}
where $\beta = k_{\mathcal{B}} \theta$ is the inverse temperature and $\tau_0 = a/u$ is a short-time cut-off. 

The form of the correlation function in Eq.~\eqref{eq:CorrelationAnyons} can be understood in a phenomenological way. Based only on the free boson theory for the Laughlin sequence, one should have $\delta = \lambda = \nu$. Nevertheless, in a real sample the scaling dimension can be renormalized~\cite{Braggio12}. In contrast, the contribution in Eq.~\eqref{eq:CorrelationAnyons} related to $\lambda$ arises due to the boson commutation relation and it is independent of the specific properties of the sample.  

\textit{Time-dependent charge current:-} The goal is to investigate the braiding of anyons by looking at transport properties, which are computed in terms of the correlation functions presented above. We start by considering the time-dependent current, which is defined as (we choose $e>0$)
\begin{equation}
 I(t) = -e u \rho\left(D,t\right),
 \label{eq:CurrentDefinition}
\end{equation}
where $\rho(x,t) = -\sqrt{\nu}/\left(2\pi\right) \partial_x \phi(x,t)$ is the total charge density expressed in terms of boson fields~\cite{Wen95}, evolving with respect to the full Hamiltonian, and $D$ is the position of the detector (see Fig.~\ref{fig:scheme}). Since the tunneling is tuned to the weak-tunneling regime, we calculate the current perturbatively to the lowest order in the tunneling amplitude $\Lambda$. Unlike the transport calculations for a standard quantum point contact (QPC) involving two edges, where the lowest contributing order is second order, the first-order term is already non-vanishing in this case.

\begin{figure}[t]
\hspace{-10mm}\centering
\includegraphics[width=0.5\textwidth]{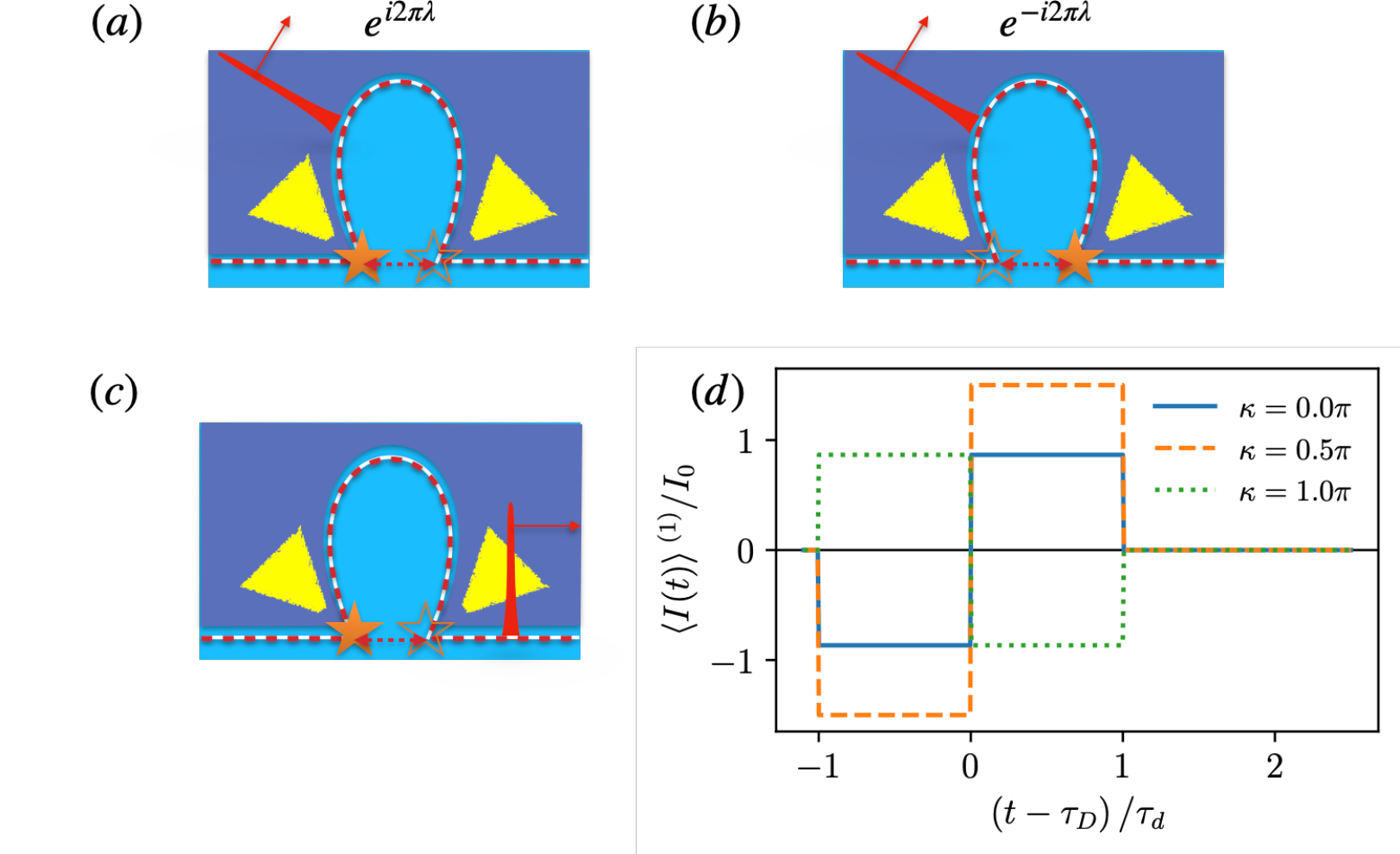}
\caption{(a) A single injected anyon is passing through the loop while a spontaneous quasiparticle/quasihole pair (orange stars) is created, thus giving rise to a non-trivial braiding phase $e^{i 2\pi \lambda}$. (b) Similar process with creation of a spontaneous quasihole/quasiparticle pair: the associated braiding phase is the complex conjugate of the one in panel (a). (c) The single injected anyon has already passed through the loop while the pair is created: no braiding phase is associated with this process. (d) The total charge current as a function of $\left(t-\tau_D\right)/\tau_d$ for different values of the phase $\kappa$ {at $\nu=1/3$}. Through time, the value of charge is changing following the dynamics described by the processes in panels (a)-(c). In particular, the interference between the two processes in (a) and (b) gives rise to a finite current.}
\label{fig:OneAnyon}
\end{figure}

Before providing the explicit expression for the current, it is instructive to describe the physical picture for the tunneling at the QPC. Spontaneous creation of quasiparticle-quasihole pairs occurs continuously at the junction, 
with the process $P_1$ (in which a quasiparticle is created at $x = -d/2$ and a quasihole at $x = +d/2$) and its reverse process $P_2$.
These processes create two contributions for the current along the edge: one at $x = -d/2$ and another at $x = d/2$. We denote these contributions as the tunneling currents $I_T^-$ and $I_T^+$, respectively. 
These currents satisfy $I_T^+ (t) = -I_T^-(t)$, reflecting the opposite charges of the spontaneously created anyons in both the $P_1$ and $P_2$ processes.

In the absence of incoming anyonic excitations, the process $P_1$ and its reverse process $P_2$ are characterized by phases $e^{-i\kappa}$ and $e^{i\kappa}$, respectively (see Eq.~(\ref{eq:HamTun})). As a result, the tunneling currents become $I_T^{\pm} = \mp \sin(\kappa)$, thus generating no current along the edge beyond the droplet.

Now, let us consider the case where a single anyon is incoming on the junction, as illustrated in Fig.~\ref{fig:OneAnyon}. If a spontaneous quasiparticle-quasihole pair is emitted during the time interval $\tau_d = d/u$, where the incoming anyon is within the loop, the phases of the $P_1$ and $P_2$ processes are modified by a factor of $\pm 2\pi\lambda$, due to the anyonic exchange statistics. As the currents $I_T^{\pm}$ become time-dependent, and as $I_T^+(t)$ and $I_T^-(t)$ are emitted at different positions along the edge, this creates a non-zero current beyond the droplet.

While all the details for the calculations are carried out in Ref.~\cite{Ronetti25}, here we present the final result for the time-dependent charge current, whose contribution at first order in the tunneling amplitude reads $\left\langle I(t) \right\rangle^{(1)} = I^+_T\left(t-\tau_d/2\right)+ I_T^-\left(t+\tau_d/2\right)$, with 
\begin{equation}
	I^{\pm}_T\left(t\right) = \mp I_0\left(\delta,\theta\right) \sin\left[2\pi \lambda N\left(t\right) +\kappa\right]\label{eq:TimeDependentChargeCurrentFirstOrder}
\end{equation}
where $I_0\left(\delta,\theta\right) \equiv e^*\Lambda e^{\delta \mathcal{G}\left(d/u\right)}/\left(\pi a\right)$ is the current amplitude, $N(t) = \int^{t-\tau_D+\tau_d/2}_{t-\tau_D-\tau_d/2} \! d\tau \; \delta(\tau) = 0 \mbox{ or } 1$ if the incoming anyon is outside or inside the loop at time $t - \tau_D$, with $\tau_D = D/u$ the time needed to reach the detector. Manifestly, the current at first order is finite only if $I^+_T\left(t-\tau_d/2\right)+ I_T^-\left(t+\tau_d/2\right)\ne 0$. Eq.~\eqref{eq:TimeDependentChargeCurrentFirstOrder} represents one of the main results of this Letter. We remark that for the integer filling factor $\nu = 1$, when fermions are injected, the current at first order vanishes, as proven in Ref.~\cite{Ronetti25}. More importantly, it is worth noting that the result is non-zero even at first order in the tunneling amplitude.

The time-dependent charge current is shown in Fig.~\ref{fig:OneAnyon}(d). Each of the two steps, with opposite signs, corresponds to the tunneling currents $I_T^\pm(t)$. These steps are separated by a time delay $\tau_d = d/u$, which reflects the time it takes to travel through the loop.

In contrast with a conventional QPC, here it is the superposition of equal-time processes $P_1$ and $P_2$, with different statistical phases, that gives rise to a finite charge current. Therefore, no integral over time is required and the current depends only on the creation amplitude for one pair at the QPC at time $t=0$, i.e. $e^{\delta \mathcal{G}\left(d/u\right)}$. As a result, the effects of both the scaling dimension $\delta$ and the temperature $\theta$ factorize with respect to the contribution of the statistical angle $\pi\lambda$. This crucial result arises from the fact that the current is non-vanishing even at first order in the tunneling amplitude.

\begin{figure}[t]
\centering
\includegraphics[width=0.49\textwidth,angle=0]{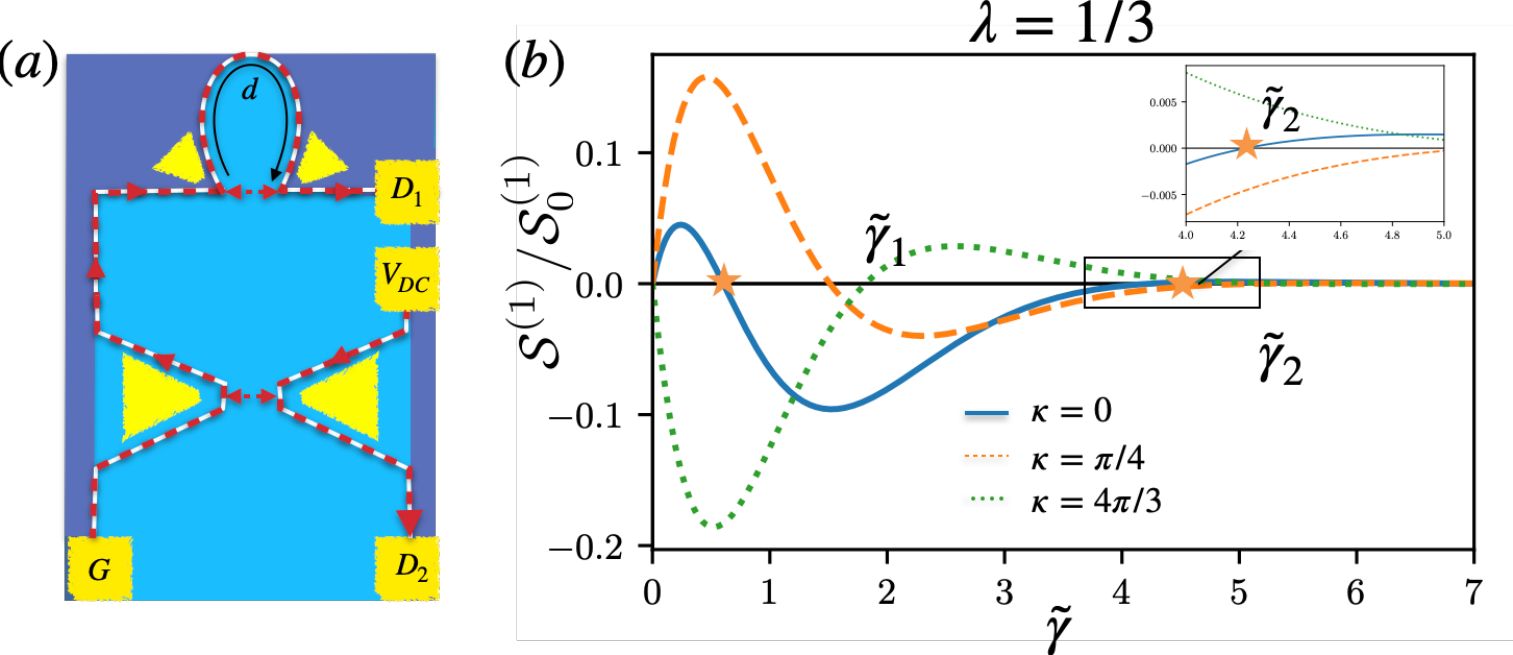}
\caption{(a) Setup for the measurement of the statistical angle using the $\Omega$ junction with a dilute stream of anyons. A QPC is placed upstream with respect to the droplet and tuned to the weak-backscattering regime. In this way, a stream of anyons distributed according to Poissonian statistics impinges at the $\Omega-$junction. The current correlations between the detectors $D_1$ and $D_2$ are measured to extract the statistical angle $\lambda$. (b) The finite-frequency noise $\mathcal{S}^{(1)}/\mathcal{S}^{(1)}_0$ as a function of $\tilde{\gamma}$ for $\lambda = 1/3 $ and various values of $\kappa$. From the position of the zeros $\tilde{\gamma}_1$ and $\tilde{\gamma}_2$, one can obtain the statistical angle $\lambda$ (see Eq.~\eqref{eq:Lambda}). The inset shows a zoomed-in view of the position of the second zero. }
\label{fig:SetupCollider}
\end{figure}

\textit{Dilute stream of anyons:-} While the physics presented for the charge current is rich and interesting, it is nevertheless quite complicated to envisage the measurement of time-dependent quantities generated by isolated anyon pulses. {Moreover, the statistical angle cannot be directly exctrated from the current, especially in the case $M=1$ (see Ref.~\cite{Ronetti25} for a discussion about the case $M>1$).} In the following, we focus on the case of a dilute stream of anyons injected by a source QPC placed upstream with respect to the $\Omega$ junction, as shown in Fig.~\ref{fig:SetupCollider}(a).

The source QPC is biased by a constant potential $V_{DC}$. At its output, anyons are randomly emitted and their number is distributed according to a Poisson distribution with parameter $\gamma$~\cite{Rosenow16,Bartolomei20}. It follows that the average incoming current toward the $\Omega$ junction is $e^* \gamma = \mathcal{T} V_{DC}$, where $\mathcal{T}$ is the transmission of the source QPC. In order to extract information about the braiding statistics of anyons, we compute the finite-frequency noise defined as~\cite{Chevallier10}
\begin{equation}
    \mathcal{S}\left(\omega\right) = \text{Re}\left[ \int d\left(t-t'\right) e^{i \omega \left(t-t'\right)}\left\langle \delta I_1\left(t\right)  \delta I_2\left(t'\right)\right\rangle \right],
\end{equation}
which corresponds to the cross-correlations between the excess current operators $\delta I_1 = I_1 - \left\langle I_1\right\rangle$ and $\delta I_2 = I_2 - \left\langle I_2\right\rangle$, detected at terminals $D_1$ and $D_2$, respectively (see Fig.~\ref{fig:SetupCollider}). Note that cross-correlations can take complex values at finite frequency, but we only consider the real part here as this is the only meaningful contribution from an experimental standpoint~\cite{Creux06}.

We compute this quantity at lowest order in tunneling at the $\Omega-$junction and we treat non-perturbatively the presence of the source QPC by means of the non-equilibrium bosonization formalism~\cite{Rosenow16}. In this way, one arrives at the following expression~\cite{Ronetti25}
\begin{equation}
\mathcal{S}^{(1)}\left(\omega\right) = \mathcal{S}^{(1)}_0\mathcal{C}_{\lambda}\left(\tilde{\gamma}\right)\label{eq:CrossCorrelations}
\end{equation}
where we defined $\tilde{\gamma} = \gamma \tau_d$ and
{\begin{align}
\mathcal{S}_0^{(1)}\left(\omega\right) =& 4 \frac{e^2\nu^2 \Lambda}{\pi a} e^{\delta \mathcal{G}\left(d/u\right)} \sin \left(\omega \frac{D_1-D_2}{u} \right)  \frac{ \sin^2 \left( \omega \frac{\tau_d}{2} \right)}{ \omega \frac{\tau_d}{2}},\\
\mathcal{C}_\lambda (\tilde{\gamma}) &= \sin (\pi \lambda)\tilde{\gamma} e^{-\tilde{\gamma} \left[1-\cos\left({2\pi \lambda}\right)\right]}\nonumber\\& \times\cos \left[\tilde{\gamma} \sin\left(2\pi\lambda\right)+\pi\lambda - \kappa\right],
\end{align}
with $D_1$ and $D_2$ the curvilinear abscissa of the two detectors. We note that, for fermions, $\mathcal{C}_{\lambda=1}\left(\tilde{\gamma}\right) = 0$.}

{In Fig.~\ref{fig:SetupCollider}(b), the rescaled cross-correlation noise $\mathcal{S}^{(1)}$ is shown as a function of the normalized injected anyon current $\tilde{\gamma}$. One can easily see the distinctive behaviour of $\mathcal{S}^{(1)}$ for $\lambda=1/3$, which is oscillating and decaying, compared to the fermionic case at $\lambda = 1$, which is vanishing.} 

As can be read from Eq.~\eqref{eq:CrossCorrelations}, even the finite-frequency noise depends on a non-universal constant $\kappa_{\lambda}$. However, the relative position of two consecutive zeros $\tilde{\gamma}_1$ and $\tilde{\gamma}_2$ of the sine function in Eq.~\eqref{eq:CrossCorrelations} is always independent of $\tilde{\kappa}_{\lambda}$. We give an example for $\kappa = \pi/4$ in Fig.~\ref{fig:SetupCollider}(c). In this case, by inverting the previous relations, the statistical angle can be directly determined by the equation
\begin{equation}
\lambda = \frac{1}{2\pi}\arcsin\left(\frac{\pi}{\tilde{\gamma}_2-\tilde{\gamma}_1}\right).\label{eq:Lambda}
\end{equation}
In general, the experimental observation of the curve shown in Fig.~\ref{fig:SetupCollider}(c) enables the extraction of the statistical angle as a fitting parameter without any prior knowledge of non-universal quantities.  More specifically, the statistical angle can be determined by measuring the noise as a function of $\gamma$, which depends simply on the potential bias and on the transmission of the source QPC. The value of $\tau_d$, which is required to obtain $\tilde{\gamma}$ via $\tilde{\gamma}= \gamma \tau_d$, can then be determined together with $\lambda$ by combining Eq.(13) and the fact the the exponential decay of the noise is given by $\tilde{\gamma}(1-\cos(2 \pi \lambda))$.

The range of relevant values of $\tilde{\gamma}$, as plotted in Fig.~\ref{fig:SetupCollider}, should be accessible experimentally. At filling factor $\nu = 1/3$, for typical samples with edge velocity \( u \sim 10^4~\unit{\meter\per\second} \) and maximum injected current \( I_\text{inj} = e^* \gamma \sim 1~\unit{\nano\ampere} \), this corresponds to a loop of length \( d \sim 3~\unit{\micro\meter} \), which is well within reach of current nanofabrication techniques. {This loop size balances the competing requirements of suppressing edge reconstruction and minimizing thermal dephasing, in agreement with experimental observations of coherent quantum interference over similar length scales in the fractional quantum Hall regime~\cite{Nakamura20}. }

{Another important point regarding experimental feasibility concerns the presence of disorder. While disorder in the $\Omega$-loop region may induce sample-dependent phase shifts, it does not obscure the topological signal: the statistical angle is encoded in a universal contribution that remains robust against non-universal prefactors such as $\kappa$, which are fixed for a given sample realization.}

It is also important to stress that the actual frequency dependence of the cross-correlation noise is irrelevant for the determination of the statistical angle, so that a measurement carried out at a single, low but finite frequency would be sufficient. The results in Eqs.~\eqref{eq:CrossCorrelations} and~\eqref{eq:Lambda}, and the related discussion, represent the second key finding of our Letter.

\textit{Conclusions:-} In this Letter, we proposed an experimental protocol to detect directly the statistical angle of anyons in the fractional quantum Hall regime. Our setup consists of a single edge of an Abelian fractional quantum Hall liquid bent to form an $\Omega-$shaped junction where injected anyons can impinge at. Firstly, we presented the main physics associated with this setup by focusing on the case of an isolated delta-like anyon pulse. We showed that the time-dependent tunneling current is non-vanishing at first order in the tunneling amplitude and that the effects of temperature and scaling dimension factorize with respect to the contribution of the anyon statistics. In the presence of a dilute stream of fractional quasi-particle emitted by a source QPC, we computed the finite-frequency cross-correlations noise, and we showed how the statistical angle $\lambda$ can be accessed directly. Our results introduce a new configuration where anyon statistics can be tested with crucial advantage compared to the existing ones. We believe that an important extension for our calculations is to consider other Abelian filling factors or non-Abelian quantum Hall states, which will allow a unique exploration of the more complex dependence on $\lambda$ and $\delta$.

{Another promising direction for future work is to investigate how our results are modified by nonlinear corrections to the chiral Luttinger liquid theory, such as those discussed in Refs.~\cite{Fern18, Nardin23, Monteiro24}, which go beyond the linear edge dispersion assumed in our model.
}

\acknowledgments{M. H. would like to thank Victor Bastidas for fruitful discussions at the very first stage of this work. This work was carried out in the framework of the project ``ANY-HALL" (ANR Grant No.
ANR-21-CE30-0064-03). It received support from the French government under the France 2030 investment plan, as part of the Initiative d’Excellence d’Aix-Marseille Université A*MIDEX. We acknowledge support from the institutes IPhU (AMX-19-IET008) and AMUtech (AMX-19-IET-01X). This work has benefited from State aid managed by the Agence Nationale de la Recherche under the France 2030 programme, reference ``ANR-22-PETQ-0012".
This French-Japanese collaboration is supported by the CNRS International Research Project ``Excitations in Correlated Electron Systems driven in the GigaHertz range" (ESEC).}

\end{document}